\documentclass[a4paper]{jpconf}
\usepackage{graphicx,amsmath,amssymb}
\pdfoutput=1

\newcommand{\be}{\begin{equation}}
\newcommand{\ee}{\end{equation}}
\newcommand{\bea}{\begin{eqnarray}}
\newcommand{\eea}{\end{eqnarray}}

\newcommand{\comment}[1]{}

\newcommand{\sm}[1]{{$#1$}}
\newcommand{\seq}[1]{\begin{equation}
   #1
\end{equation}}
\newcommand{\seqa}[1]{\begin{eqnarray}
  #1
\end{eqnarray}}

\begin{document}

\begin{flushright}
{\it CP$^\mathit{3}$-Origins-2010-42\\
 CCTP-2010-14 }
\end{flushright}

\title{Electroweak phase transition in technicolor}

\author{Matti J\"arvinen\footnote{Present address: Crete Center for Theoretical Physics, Department of Physics, University of Crete, 71003 Heraklion, Greece}}

\address{CP$^3$-Origins, Campusvej 55, 5230 Odense M, Denmark}

\ead{mjarvine@physics.uoc.gr}

\begin{abstract}

Several phenomenologically viable walking technicolor models have been proposed recently. I demonstrate that these models can have first order electroweak phase transitions, which are sufficiently strong for electroweak baryogenesis. Strong dynamics can also lead to several separate transitions at the electroweak scale, with the possibility of a temporary restoration and an extra breaking of the electroweak symmetry. First order phase transitions will produce gravitational waves, which may be detectable at future experiments.

\end{abstract}

\section{Introduction}

 Progress in the understanding of the phase diagram of generic asymptotically free gauge theories has led to  a renewed interest in technicolor models (for a recent review see \cite{Sannino:2009za}).  {}Explicit examples of technicolor models, not in conflict with electroweak precision tests, have been put forward in \cite{Sannino:2004qp,Dietrich:2005jn,Dietrich:2006cm,Ryttov:2008xe}. 

I review the recent work \cite{Cline:2008hr,Jarvinen:2009pk,Jarvinen:2009wr,Jarvinen:2009mh} 
on the finite temperature electroweak phase transition in technicolor.  The standard model has a cross-over electroweak transition \cite{Kajantie:1995kf}. It is interesting to see if first order phase transitions are possible in technicolor \cite{Kikukawa:2007zk} or other \cite{CQW,DR,CM,LR} extensions of the standard model, because this would remove an obstacle for realizing electroweak baryogenesis (see, e.g., \cite{Cline:2006ts}). 

This work concentrates on Minimal Walking Technicolor (MWT) \cite{Sannino:2004qp} and Ultra Minimal Technicolor (UMT) \cite{Ryttov:2008xe}. These two models can be seen as templates for two qualitatively different cases of generic strongly interacting extensions of the standard model. I use effective field theory to describe the strong dynamics of the strongly interacting technicolor sector. The parameter space of the effective theory covers a variety of different possible strong dynamics. The fact that UMT has two matter sectors and therefore in general two chiral phase transitions makes it essentially different from MWT. 

In MWT, the standard model Higgs is replaced by two (Dirac) techniquarks which transform in the {adjoint} representation of the technicolor gauge group $SU(2)_{\rm TC}$. MWT as such suffers from the Witten topological anomaly \cite{Witten:1982fp}, which is cured by adding an extra generation of massive leptons. 
UMT has two sectors of matter transforming under the fundamental and adjoint representations of  \sm{SU(2)_{\rm TC}}, leading typically to two chiral phase transitions. The fundamental sector contains two Dirac techniquarks, and the adjoint sector has two Weyl techniquarks.

\section{Tools}

\subsection{Effective theory}

The relevant degrees of freedom in the technicolor theory at the electroweak scale are the (lightest) composite technihadrons. Their dynamics cannot be solved from the underlying strongly interacting theory, but it can be described by using an effective field theory, which implements the chiral symmetry breaking pattern (\sm{SU(4)\to SO(4)} for MWT) \cite{Appelquist:1999dq}. In this study I include the lightest scalar states.

For MWT, the simplest effective Lagrangian of the scalar sector reads
\seq{\label{eq:LMWT}{\cal L}_{\rm eff} = \frac{1}{2}{\rm Tr}\left[D_{\mu}M D^{\mu}M^{\dagger}\right] - {\cal V}(M) + {\cal L}_{\rm ETC}}
where
\seqa{
{\cal V}(M) & = & - \frac{m^2}{2}{\rm Tr}[MM^{\dagger}] +\frac{\lambda}{4} {\rm Tr}\left[MM^{\dagger} \right]^2 
+ \lambda^\prime {\rm Tr}\left[M M^{\dagger} M M^{\dagger}\right] \nonumber \\
& - & 2\lambda^{\prime\prime} \left[{\rm Det}(M) + {\rm Det}(M^\dagger)\right] \ ,}
and $D_\mu$ is the electroweak covariant derivative. The matrix \sm{M} contains 20 scalar states, including the pions which are eaten by the \sm{Z} and \sm{W} bosons in the electroweak transition, a set of extra Goldstone modes, and a composite Higgs. Consistency requires us to include also the pseudoscalar partner $\Theta$ of the Higgs and a scalar partner triplet $A$ of the eaten pions. The masses of these states may be close to the electroweak scale, and therefore they play an important role in the electroweak phase transition. I also need to add masses to the extra Goldstone modes.  These masses are contained in the term ${\cal L}_{\rm ETC}$ above. They may arise through the coupling to the electroweak sector \cite{Dietrich:2009ix} and/or through extended technicolor interactions. The couplings $\lambda$, $\lambda^\prime$ and $\lambda^{\prime\prime}$ are free parameters in this study, and they can be traded for a set of masses of the composite particles. In the future, these parameters may be constrained by lattice studies or by experimental data.

The chiral symmetry of UMT is $SU(4)\times SU(2) \times U(1)$, including the separate $SU(4)$ and $SU(2)$ transformations of the two sectors of matter and one anomaly free $U(1)$ transformation. The symmetry is expected to break spontaneously to $Sp(4) \times SO(2) \times Z_2$. The dynamics of the two matter sectors of UMT are described by Lagrangians analogous to \eqref{eq:LMWT}. The chiral symmetry also allows terms which mix the two sectors \cite{Ryttov:2008xe,Jarvinen:2009pk}. At the level of elementary particles, these terms can be understood to arise from technigluon exchange between the two types of matter.

\subsection{Effective potential} 

I study the phase transition by using the standard effective potential:
\seq{V(\phi,T) = V_{\rm tree}(\phi) + V_{\rm 1-loop}(\phi,T=0)+ V_{\rm 1-loop}(\phi,T)}
where \sm{\phi} is the vacuum expectation value of the composite Higgs. I use the effective Lagrangian described above to describe the dynamics of technihadrons. For UMT the potential depends on two Higgs VEVs, corresponding to its two sectors.
The one-loop temperature dependent term includes ring resummation \cite{Arnold:1992rz}. The top and the electroweak gauge bosons are the only relevant standard model particles. I also use the method of \cite{Cline:1996mga} to interpolate between the high and low temperature asymptotics of the temperature-dependent one-loop term.

At high temperatures the effective potential is minimized at $\phi=0$, and the electroweak symmetry is intact. First order phase transition is possible if the effective potential admits another minimum at nonzero $\phi=\phi_{\rm br}$ at low temperatures, separated from the symmetric one by a potential barrier. Below the critical temperature $T_c$, the symmetric vacuum becomes unstable: $V(0,T)>V(\phi_{\rm br},T)$. First order phase transition proceeds via bubble nucleation at some temperature $T_*$ ($T_*<T_c$).
The bubble profile for the above effective potential can be solved numerically, and it can be used to estimate the bubble nucleation rate.
$T_*$ is then defined as the temperature where the nucleation rate per Hubble volume and time becomes equal to one. The parameter which characterizes the strength of the transition is $\phi_*/T_*$, where $\phi_*$ is the Higgs expectation value in the broken phase vacuum at the nucleation temperature. Electroweak baryogenesis requires 
\begin{equation}
 \frac{\phi_*}{T_*} \gtrsim 1 \ .
\end{equation}

\section{Results}

\subsection{Strength of the electroweak transition in MWT}

\begin{figure} \label{fig:phTMWT}
  \includegraphics[width=0.4\hsize]{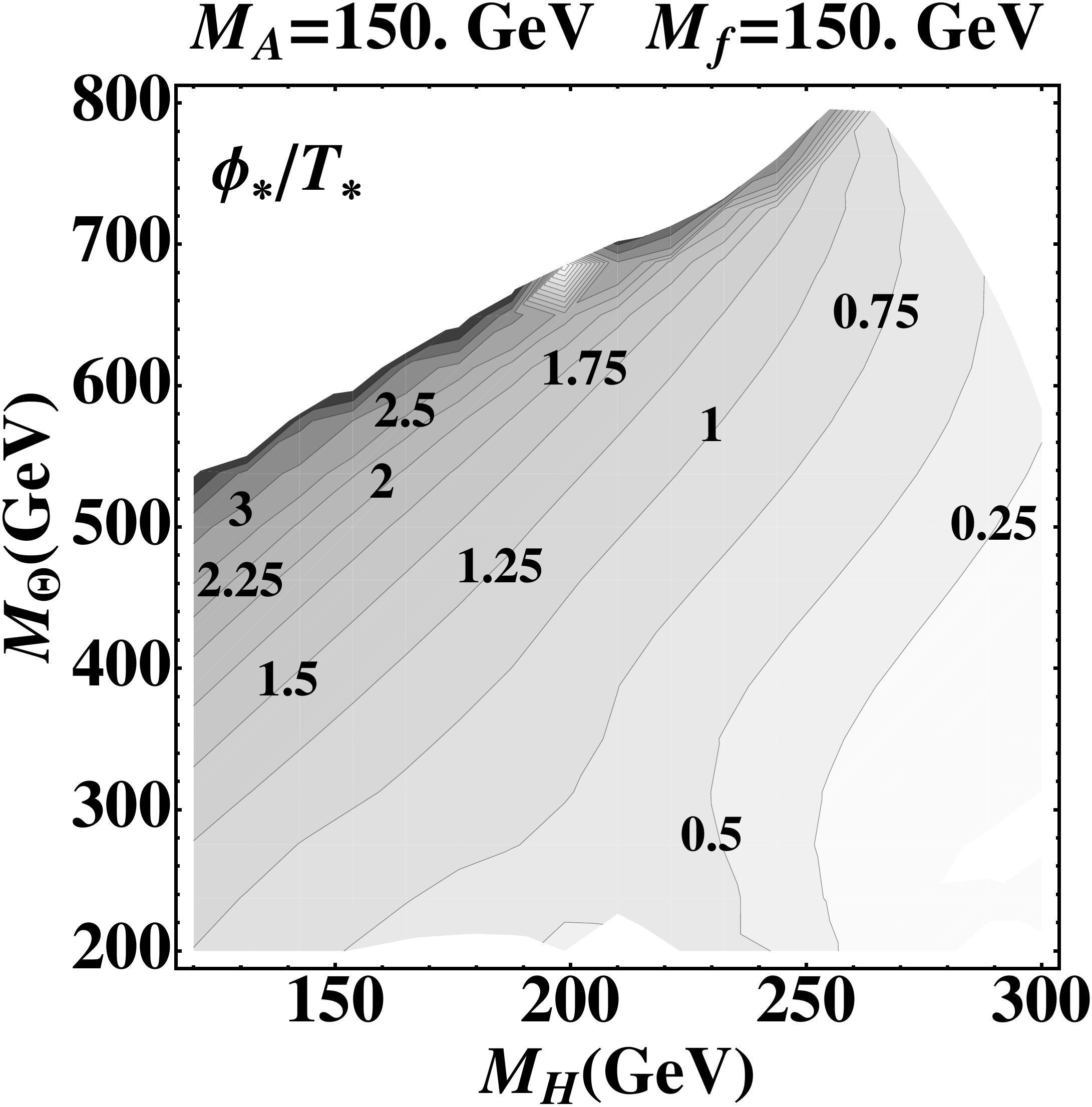}\includegraphics[width=0.4\hsize]{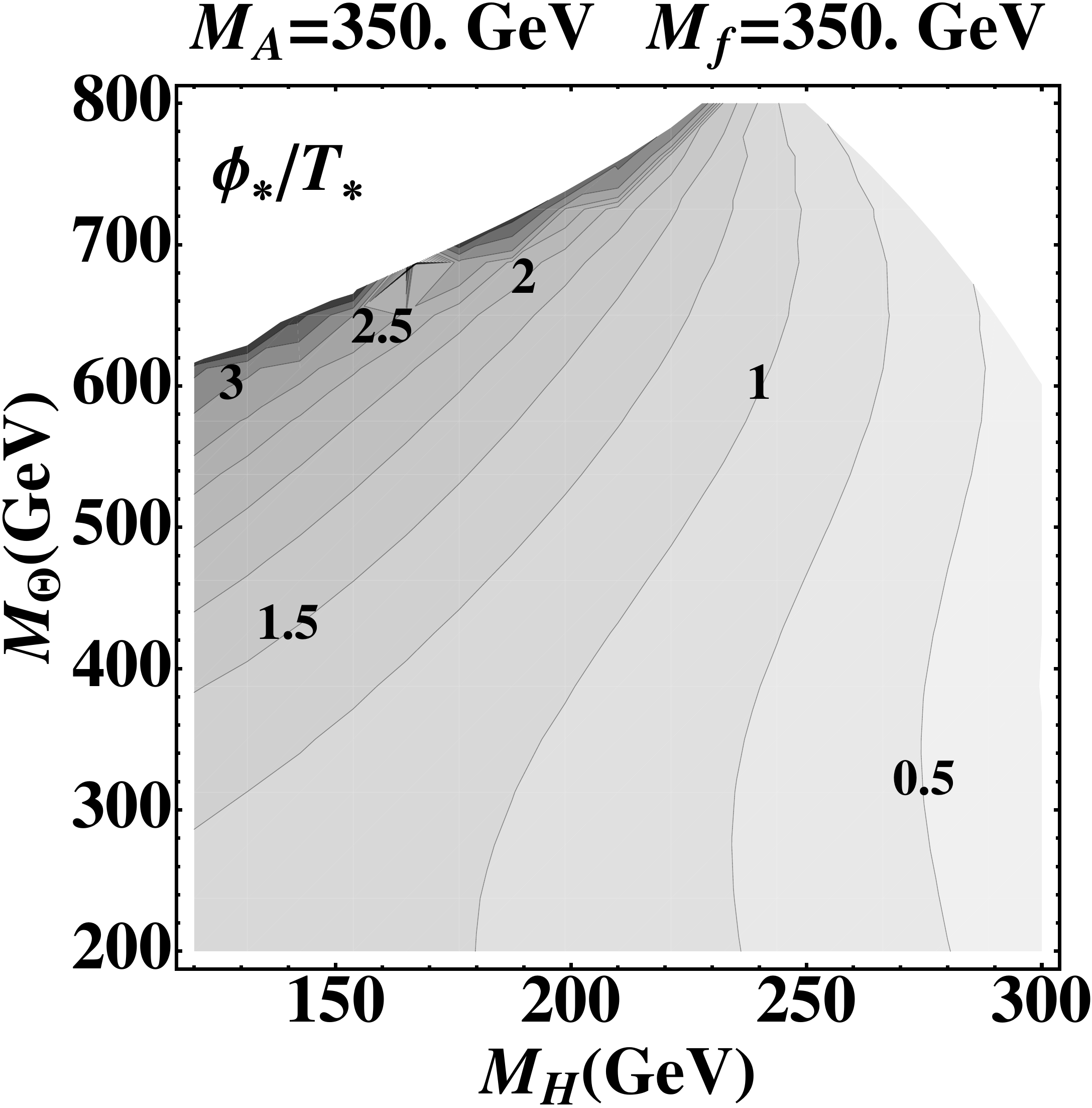}
  \caption{The strength \sm{\phi_*/T_*} of the electroweak transition in MWT for various parameter values.}
\end{figure}

The plot the ratio \sm{\phi_*/T_*} for MWT in Fig.~\ref{fig:phTMWT} \cite{Cline:2008hr,Jarvinen:2009mh}. In these plots I assumed that the masses of the extra Goldstone modes of MWT, which arise from ${\cal L}_{\rm ETC}$, are much higher than the electroweak scale so that the contributions from these Goldstones can be neglected. On the horizontal axis I have the mass of the composite Higgs, and the vertical axis has the mass of the pseudoscalar partner $\Theta$ of the Higgs. The additional parameters are the mass of the scalar partners $A$ of the eaten pions, and the mass $M_f$ of the fourth family leptons. The transition is strong enough for electroweak baryogenesis (\sm{\phi_*/T_*\gtrsim 1}) in a vast region of parameter space.

\subsection{Strength of the electroweak transition in UMT}

\begin{figure}  \label{fig:phTUMT}
  \includegraphics[width=0.4\hsize]{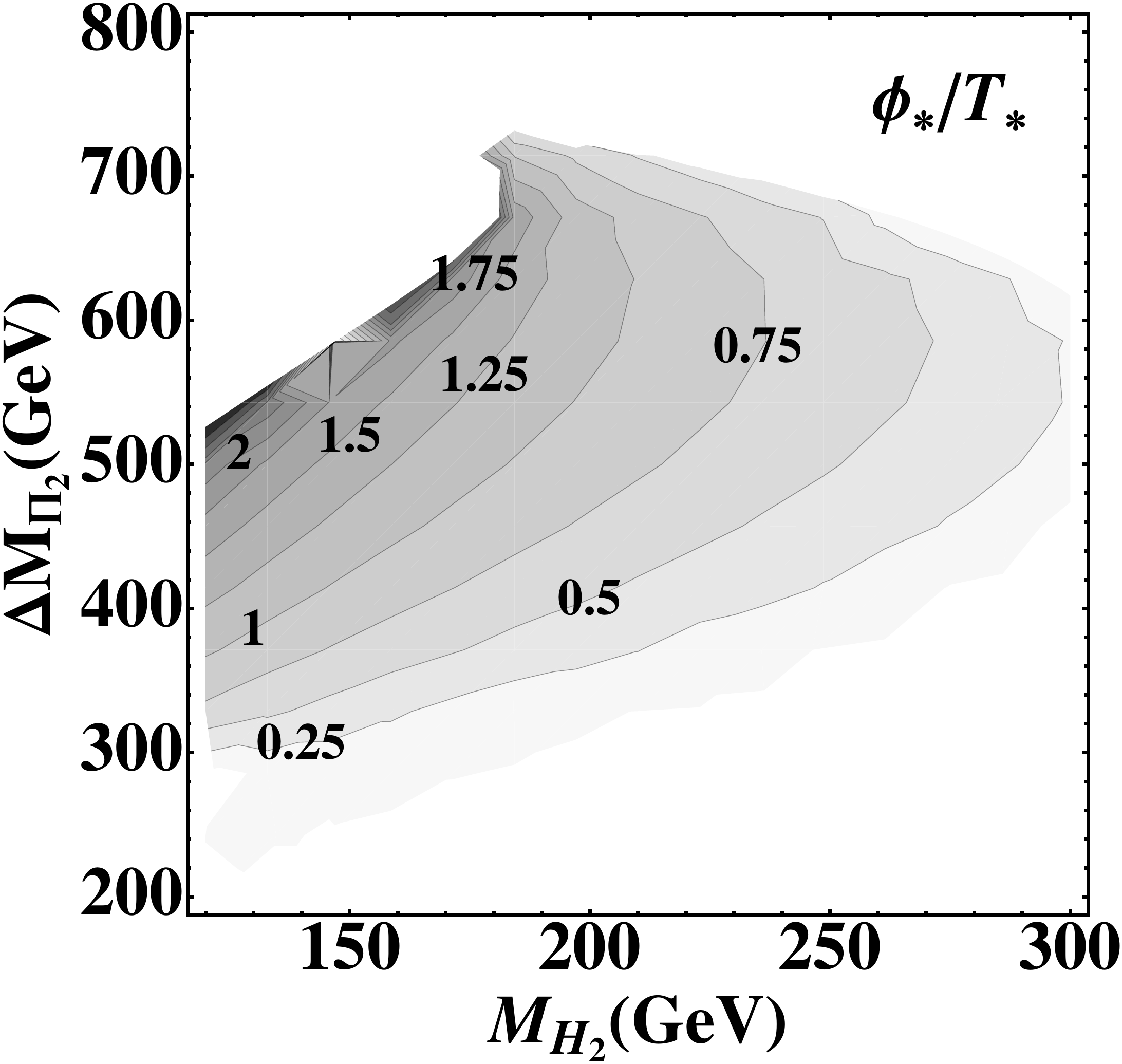}\includegraphics[width=0.4\hsize]{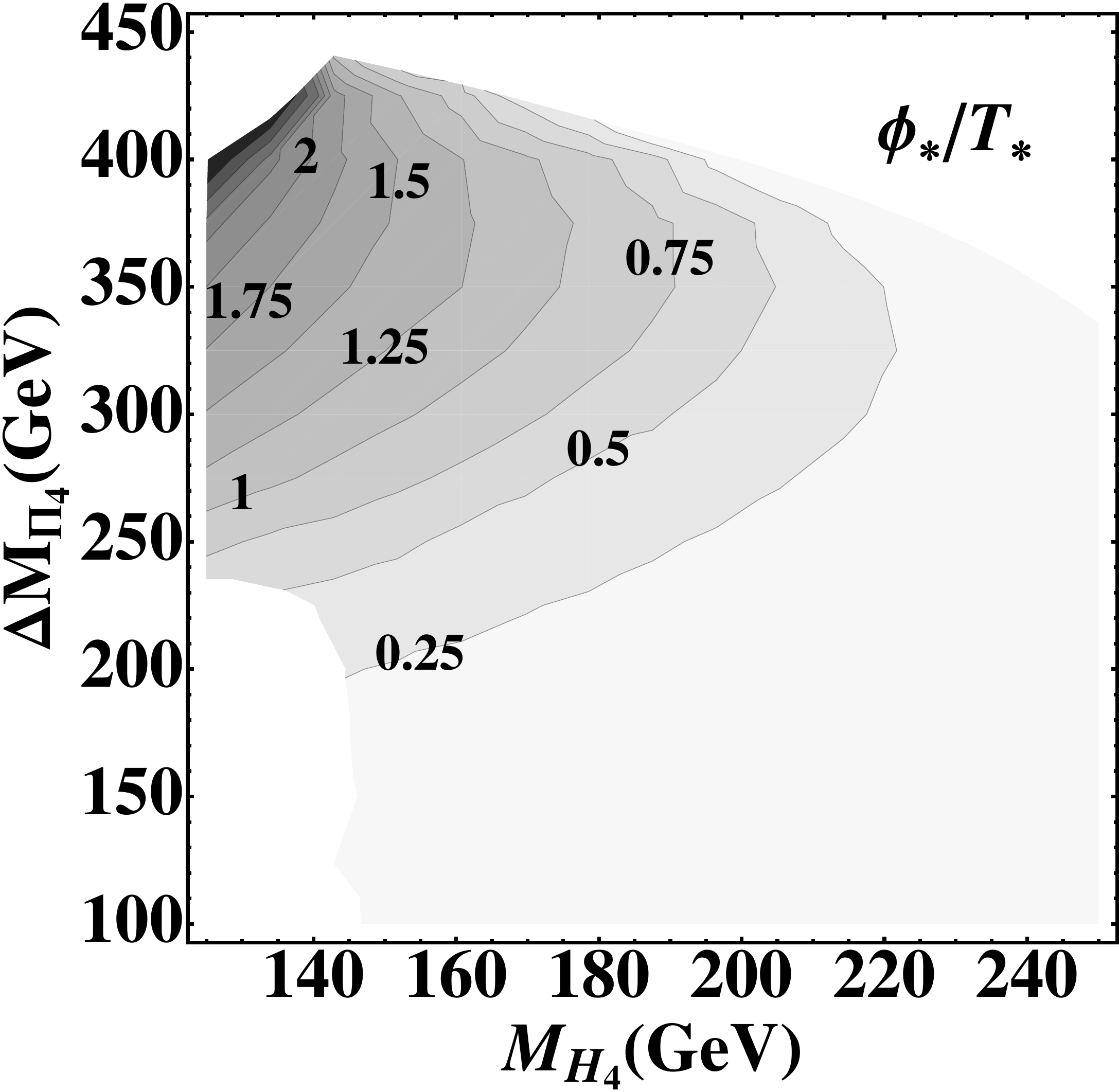}
  \caption{The strength \sm{\phi_*/T_*} of the phase transitions of the two sectors in UMT. Left: transition of matter transforming under the adjoint of $SU(2)_{\rm TC}$. Right: transition of matter transforming under the fundamental of $SU(2)_{\rm TC}$.}
\end{figure}

The parameter \sm{\phi_*/T_*} is shown in Fig.~\ref{fig:phTUMT} for both the transitions of UMT in the case where coupling between two sectors of matter is absent \cite{Jarvinen:2009pk,Jarvinen:2009mh}. The horizontal axes have the corresponding composite Higgs masses, and the vertical axes have the masses of the scalar partners of the Goldstone bosons. The right hand plot presents the strength of the transition related to the techniquarks which transform under the fundamental representation of $SU(2)_{\rm TC}$ and are gauged under the electroweak. Therefore, this transition triggers the electroweak phase transition. The effect of mixing between the two matter sectors on the strength of the transitions was analyzed in \cite{Jarvinen:2009pk}. It was typically seen to weaken the transitions.

\subsection{Extra electroweak transition in UMT}
 
\begin{figure} \label{fig:UMTpd}
  \includegraphics[width=0.33\hsize]{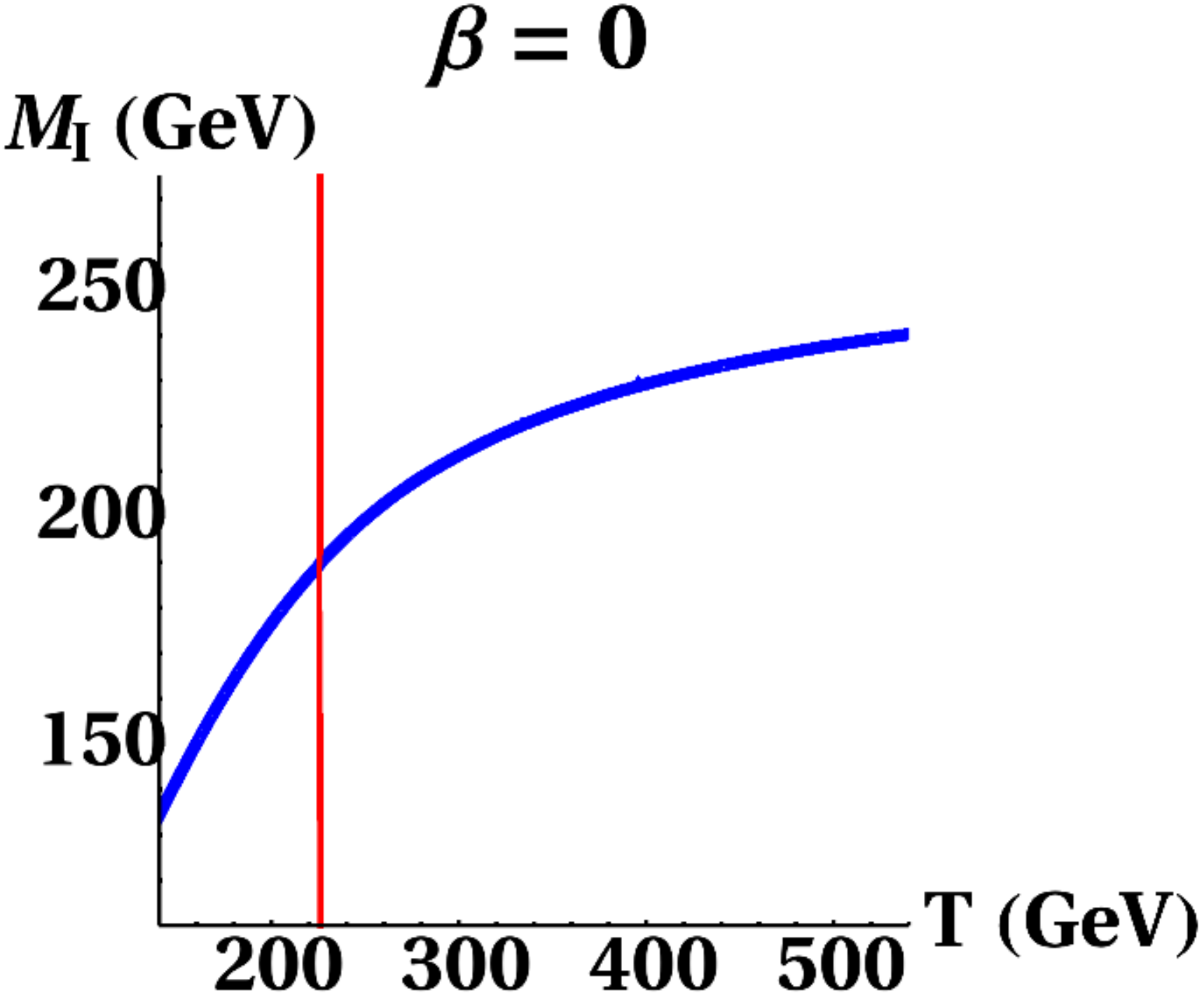}\includegraphics[width=0.33\hsize]{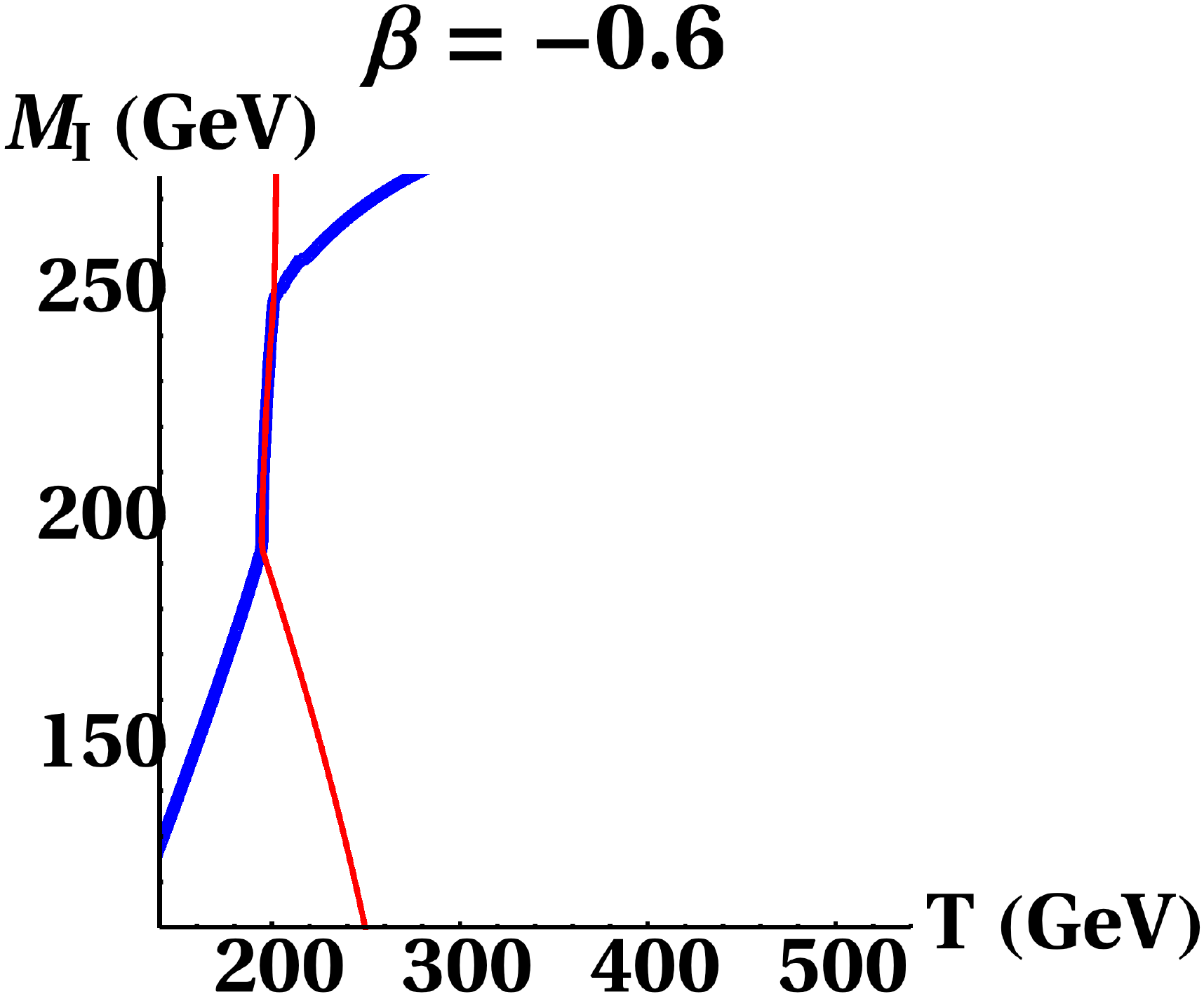}\includegraphics[width=0.33\hsize]{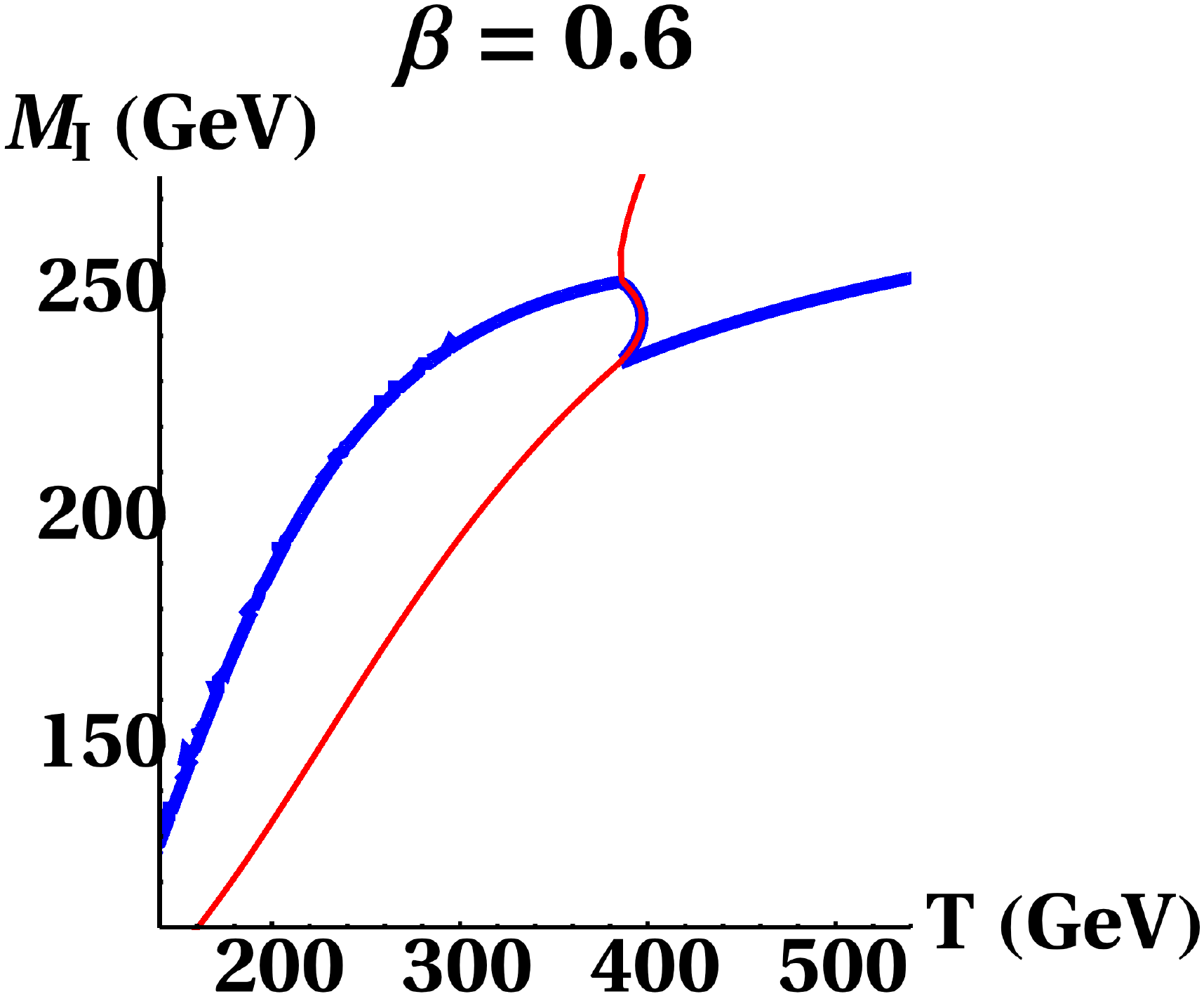}
  \caption{Phase diagrams for UMT in the $T$ -- $M_{H_4}$ plane. There is the possibility to have a simultaneous transition of both the sectors (middle), and also for an extra electroweak phase transition (right).}
\end{figure}

The two matter sectors of UMT interact strongly, which may lead to nontrivially structured phase transitions \cite{Jarvinen:2009wr,Jarvinen:2009pk}. Three qualitatively different phase diagrams on the $T$ -- $M_I$ plane, where $M_I = M_{H_4}$ is the Higgs mass of the sector charged under electroweak, are shown in Fig.~\ref{fig:UMTpd}. The thick blue line is the transition of the matter which transforms under the fundamental of $SU(2)_{\rm TC}$, i.e., the electroweak transition, and the thin red line is the transition of the sector containing matter in the adjoint of $SU(2)_{\rm TC}$.  Since the Higgs mass \sm{M_I} is (unknown but) fixed by the underlying theory, the diagrams are probed along horizontal lines from right to left as the universe cools down.

The left plot shows a phase diagram in the absence of interactions between the two matter sectors (the mixing angle $\beta$ between the two Higgses is zero). In the other plots the interactions are turned on ($\beta \ne 0$). 
The middle plot shows a case
(for well chosen parameter values \cite{Jarvinen:2009wr}) where it is possible to have a simultaneous transition of both the sectors (blue and red transitions overlap). The right hand plot shows an even more interesting case where electroweak symmetry (the blue line) is broken {twice} and restored once for $M_I$ around 240~GeV, which potentially enhances baryogenesis.

\subsection{Production of gravitational waves}

\begin{figure} \label{fig:gw}
  \includegraphics[width=0.5\hsize]{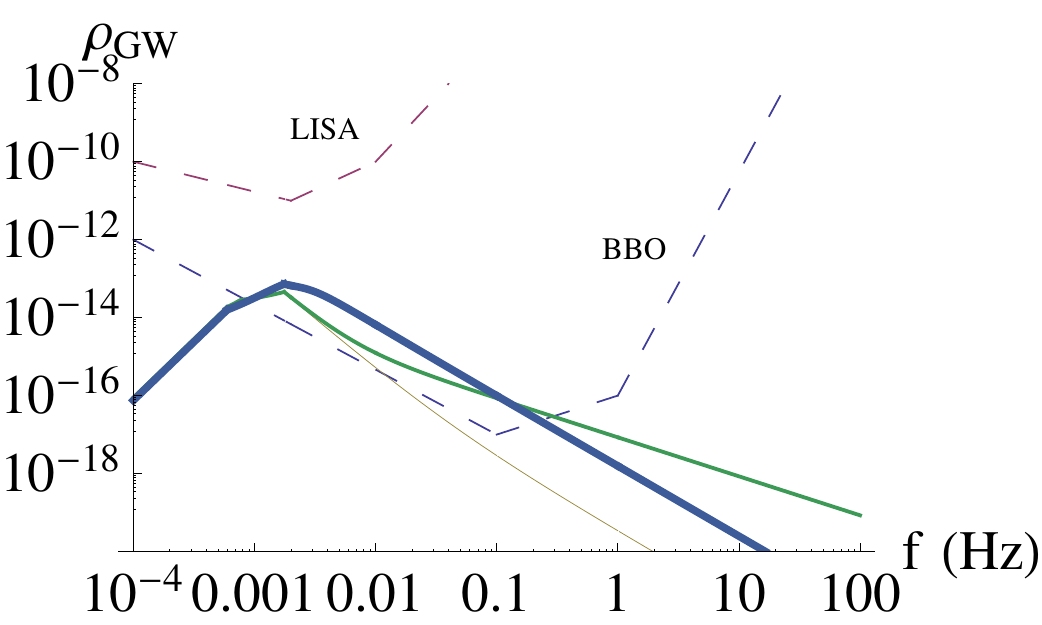}\includegraphics[width=0.5\hsize]{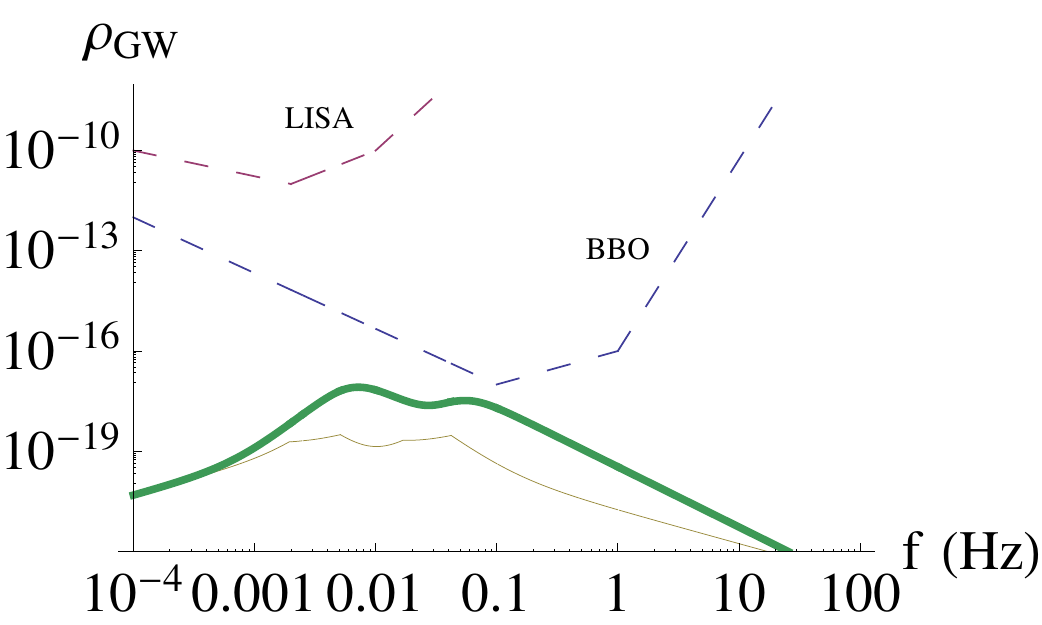}
  \caption{The energy density of the gravitational waves produced by the phase transitions for specific choices of parameter values. Left: the spectrum for MWT compared to the expected reach of LISA and BBO. Right: estimates for UMT.}
\end{figure}

The bubble profile can also be used to estimate the strength of the gravitational waves produced at the phase transition(s). I plot the spectrum of the produced waves for rather optimal, but not fine-tuned choice of parameters both for MWT and for UMT in Fig.~\ref{fig:gw} \cite{Jarvinen:2009mh}. The various solid lines follow different estimates for the production of gravitational waves from bubble collisions and from the turbulent motion of the plasma \cite{Kamionkowski:1993fg,Gogoberidze:2007an}. The dashed lines are the estimated sensitivities of the LISA \cite{Bender} and the Big Bang Observer \cite{Harry:2006fi} experiments.
For MWT (left hand plot) the produced waves can possibly be detected at the Big Bang Observer, while for UMT (right hand plot) the signal appears to be too weak. Notice however that the energy spectrum of UMT has several distinct peaks produced by the transitions of the two matter sectors and by the different wave production mechanisms. This demonstrates the fact that the structure of the phase transition may be probed through the observed spectrum of gravitational waves.

\section{Conclusion}

I analyzed the electroweak phase transition in two technicolor models, MWT and UMT, by using effective field theories. These two examples demonstrate that strong first order electroweak phase transition is possible in strongly interacting extensions of the standard model. Therefore, such extensions have the potential to drive electroweak baryogenesis. I pointed out that strong interactions at the electroweak scale may lead to nontrivial phase structures, and possibly to extra electroweak phase transitions. First order phase transitions also produce gravitational waves, which can potentially be detected at future experiments.

After this talk was presented, I finished a related work in collaboration with F.~Sannino \cite{Jarvinen:2010ks}, which analyzes chiral phase transitions in technicolor theories by using chiral perturbation theory. This approach suggests that the critical temperatures specifically in MWT and UMT are higher than the electroweak scale, so that $\phi_*/T_* < 1$ or the transitions are of second order. In any case, partially gauged technicolor models \cite{Dietrich:2005jn,Dietrich:2006cm} remain as good candidates for producing strong first order electroweak phase transitions. In these models the transitions are expected to be strengthened thanks to the larger number of techniflavors.

\ack

This work was done in collaboration with J.~M.~Cline, C.~Kouvaris, T.~A.~Ryttov, and F.~Sannino. I was partially supported by the Marie Curie Excellence Grant under contract MEXT-CT-2004-013510, by the Villum Kann Rasmussen foundation, by Regional Potential program of the E.U. FP7-REGPOT-2008-1:  CreteHEPCosmo-228644, and by Marie Curie contract PIRG06-GA-2009-256487.

\end{document}